\documentclass{aa}

\usepackage{graphicx}
\usepackage{txfonts}
\usepackage{caption}
\usepackage{subcaption}
\usepackage{natbib}
\usepackage{amsmath}
\usepackage{hyperref}
\usepackage{xcolor}
\usepackage{comment}

\newcommand{\noop}[1]{}
\newcommand{\okina}{\textquoteleft}

\newcommand{\Ou}{1I/{\okina}Oumuamua}
\newcommand{\borisov}{2I/Borisov}

\newcommand{\be}{\begin{eqnarray}}
\newcommand{\ee}{\end{eqnarray}}

\newcommand{\MSun} {\mbox{$M_{\odot}$}}

\begin{document} 

   \title{
Significant interstellar object production by close stellar flybys 
   }


   \author{Susanne Pfalzner
          \inst{1,} \inst{2}
          \and
          Luis L. Aizpuru Vargas\inst{1, }\inst{2}
          \and
          Asmita Bhandare \inst{3}
          \and
          Dimitri Veras \inst{4, }\inst{5}\thanks{STFC Ernest Rutherford Fellow}
          }
             
\institute{J\"ulich Supercomputing Center, Forschungszentrum J\"ulich, 52428 J\"ulich, Germany\\
\email{s.pfalzner@fz-juelich.de}
\and 
Max Planck Institute for Radio Astronomy, Auf dem H\"ugel 69, 53121 Bonn, Germany
\and
\'Ecole normale sup\'erieure de Lyon, CRAL, UMR CNRS 5574, Universit\'e de Lyon , 46 All\'ee d’Italie, 69364 Lyon Cedex 07, France
\and
Centre for Exoplanets and Habitability, University of Warwick, Coventry CV4 7AL, UK
\and
Department of Physics, University of Warwick, Coventry CV4 7AL, UK}

\date{}

\abstract
{Within just two years, two interstellar objects (ISOs) --- \Ou~and \borisov~--- have been discovered, the first of their kind. Large quantities of planetesimals form as a by-product of planet formation. Therefore, it seems likely that ISOs are former planetesimals that became somehow unbound from their parent star. The discoveries raise the question of the dominant ISO formation process.
}
 {Here, we concentrate on planetesimals released during another star's close flybys. Such close flybys happen most frequently during the first 10 Myr of a star's life. Here, we quantify the amount of planetesimals released during close stellar flybys, their ejection velocity and likely composition. }
 {We numerically study the dependence of the effect of parabolic flybys on the mass ratio between the perturber and parent star, the periastron distance, inclination, and angle of periastron. } 
 {As expected, close prograde flybys of high-mass stars produce the most considerable amount of ISOs. Especially flybys of stars with M $>$ 5\MSun~on  trajectories closer than 250 AU can lead to more planetesimals turning into ISOs than remaining bound to the parent star. Even strongly inclined orbits do not significantly reduce the ISO production; only retrograde flybys lead to a significantly lower ISO production.  For perturbers slightly more massive than the parent star, there is a competition between ISO production and planetesimals being captured by the perturber.   Whenever ISOs are produced, they leave their parent system typically with velocities in the range of 0.5~--~2 km s$^{-1}$.  This ejection velocity is distinctly different to that of ISOs produced by planet scattering ($\sim$ 4~--~8 km $s^{-1}$) and those shed during the stellar post-main-sequence phase ($\sim$ 0.1~--~0.2 km s$^{-1}$). Using the typical disc truncation radius in various cluster environments, we find that clusters like the Orion nebula cluster are likely to produce the equivalent of 0.85 Earth-masses of ISOs per star. In contrast, compact clusters like NGC 3603 could produce up to 50 Earth-masses of ISOs per star. Our solar-system probably produced the equivalent of 2~--~3 Earth masses of ISOs, which left our solar system at a mean ejection velocity of 0.7 km s$^{-1}$.}
{Most ISOs produced by flybys should be comet-like, similar to Borisov and unlike `Oumuamua. ISOs originating from compact long-lived clusters would often show a deficiency in CO. As soon as a statistically significant sample of ISOs is discovered, the combined information of their observed velocities and composition might help in constraining the dominant production process.}
{}
\maketitle

\section{Introduction}
\label{sec:intro}

\begin{figure}[t]
\includegraphics[width=0.49\textwidth]{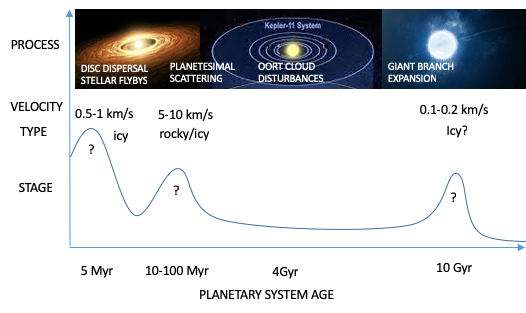}
\caption{Timeline of the ISO production mechanisms over the lifetime of  a solar-type star \citep[adapted from][]{Pfalzner:2019}. The question marks underneath the various maxima indicate that the relative amount of ISOs produced by the different mechanisms is still unknown. However, here, we can add the typical ISO ejection velocities for the various processes.}
\label{fig:phases}
\end{figure}

Star and planet formation theory has long predicted the existence of small (a few tens- to kilometre-sized) objects drifting through interstellar space \citep{McGlynn:1989}. However, only recently, the existence of such interstellar objects (ISOs) has been confirmed by the discovery of I1/`Oumuamua \citep{Bacci:2017,Williams:2017,Meech:2017a,Meech:2017b}  and I2/Borisov \citep{Borisov:2019,Jewitt:2019}. Such interstellar objects likely are former planetesimals that were ejected from the planetesimal population of another planetary system \citep[see, for example,][]{Wyatt:2008,Laughlin:2017}.  Planetesimals can become unbound from their parent star through a variety of processes \citep[][and references therein]{ISSI,Portegies:2018,Veras:2020} over a star's entire life span (see Fig.~\ref{fig:phases}). Early on in a star's life, ISOs are ejected by the system's giant planets during gas accretion \citep{Raymond:2017}. Subsequently, dynamical clearing takes place \citep{Duncan:1987, Charnoz:2003} and also dynamical instabilities among the giant planets lead to ISO ejection \citep{Raymond:2017}. Besides these more dynamic processes, planetesimals drift gently away from the Oort cloud over the entire main-sequence lifetime of a star \citep{MoroMartin:2019}. Finally, there is also an increased ejection rate towards the end of the star's giant branch phases of evolution \citep{Veras:2014}. 
Apart from internal ISO ejection triggered within the planetary system processes, ISOs are also released through external forces acting in close binary systems \citep{Jackson:2018} or star clusters. 

Here, we concentrate on the process where planetesimals are liberated from their parent disc during another star's close flyby. Each of the mechanisms mentioned above will contribute to the ISO population. However, for the moment, it remains unclear which process dominates the ISO production, as illustrated by the question marks in Fig.~\ref{fig:phases}. Here, we concentrate on the ISOs produced in close stellar flybys and two properties of the created ISO population, namely, the relative proportion of unbound particles and their ejection velocity distribution. We investigate the dependence of both properties on parameters such as periastron distance ($r_\mathrm{p}$), the relative perturber mass ($M_{21}$), the inclination ($i$), and the angle of periastron ($\omega$)  during parabolic flybys. The objective is to reach a first \emph{quantitative} understanding. It turns out that the ISOs' velocity might be the key to distinguish the most likely formation mechanism of an ISO.

The effect of stellar flybys on protoplanetary discs and planetary systems has been extensively investigated in the past \citep[for example,][]{Heller:1993,Clarke:1993,Hall:1996,Ida:2000,Breslau:2014,Bhandare:2016}. However, previous studies mainly focused on the planetesimals that remained bound to the disc or those captured by the perturber star \citep{Jilkova:2015}. They primarily investigated the effect of stellar flybys on the disc mass and the disc size. 

So far, the fate of the planetesimals becoming unbound during a close flyby has received little attention. Understanding largely remained on a qualitative level, merely stating that for prograde, coplanar encounters the relative proportion of planetesimals becoming unbound from their parent star increases for closer periastron distances and higher perturber masses \citep{Clarke:1993}. 
Here, we take the necessary next step and quantify the proportion of planetesimals that become unbound by performing an extensive parameter study, including inclined encounters.

In order to release a high number of ISOs, the flyby has to be relatively close. Such close stellar flybys happen most frequently during the first few Myr after the star has formed in a star cluster environment, particularly in the cluster cores' vicinity, where the stellar density is highest \citep{Adams:2006, Li:2015,Portegies:2019}. However, close flybys also happen during the later stages of a star's life "in the field" \citep{Li:2020}. The lower frequency of close flybys during the later stages is often balanced out by the much longer time available \citep{Spurzem:2009,Cai:2018,Pfalzner:2018,Fujii:2019, Elteren:2019}. In any case, it is the local stellar density that determines the relative frequency of close stellar flybys. This stellar density varies over many orders of magnitude between stellar groups \citep{Wolff:2007,Jaehnig:2015} and so does the average effect of the environment on the disc and planetary systems 
\citep[for example,][]{Vincke:2016, Winter:2018,Vincke:2018,Pfalzner:2020}. 

Recently, \cite{Hands:2019} took the first step towards a better understanding of the ISOs produced in cluster environments by performing simulations of clusters of young stars surrounded by discs. The discs were modelled as homogeneous rings of test particles representing the planetesimals. They modelled simultaneously the interaction dynamics of the stars and planetesimals, which is computationally expensive. Thus the number of cluster stars, the resolution of the discs and the number of simulations per model were necessarily low. They found that the number and velocity of the ejected planetesimals differ between their different cluster models. The percentage of ejected test particles typically ranging between 1 -- 4\% and the velocities between 0 and 10 km s$^{-1}$. However, the connection to actual flyby parameters, like periastron distance or mass of the perturber, could not be explored explicitly due to low-number statistics.

Here, we take a different approach by modelling the effect of a flyby on an individual disc. The flyby properties we vary across simulations (section \ref{sec:results}).  In a second step, we apply these results to our previous suite of simulations of the dynamics of stars clusters \citep{Vincke:2016,Vincke:2018} (section \ref{sec:cluster}). The method presented here has the advantage that it gives (i) detailed information on the dependence on the actual flyby parameters and (ii)~the possibility to apply the results to a wider variety of star clusters, including very massive clusters containing a few ten thousand of stars. A further advantage is that the cluster dynamics is sampled over many simulations, making the result less sensitive to variations in the initial conditions. This study allows us to provide the first quantitative results of the produced ISO population's dependencies on the cluster parameters (section \ref{sec:other_processes}). The only caveat is that this method does not determine whether the ISOs leave the cluster immediately after the flyby entirely or whether they remain in the cluster until it dissolves.

In section 4, we show that the velocities of ISOs produced by various processes differ considerably. Therefore, the ISO velocity might be a key parameter for determining the dominant ISO production mechanisms. However, care has to be taken as the original ejection velocity is not identical to the ISOs' detected velocity. The parent star's actual velocity and the processes occurring during the ISO's passage through the interstellar medium have to be taken into account. Finally, we use the results obtained in section 3 to approximate the total mass of ISOs produced by the solar system (section \ref{sec:Sun}). 

\begin{table}[t]
\caption{Simulation parameter space.}
\label{tab:set-up}
\centering
      \begin{tabular}{lrl}
  \hline \hline
        \multicolumn{2}{l}{Parameter}    & Value \\
           \hline
$M_{21}$   &  &  0.3, 0.5, 0.75, 1, 1.5, 2, 5, 10, 20, 50\\
$r_\mathrm{p}$ & [AU]      &  80, 100, 120, 140, 160, 200, 250, 350\\
$i$ & [$^{\circ}$]         & 0, 30, 60, 70, 80, 90, 120, 150, 180\\
$\omega$ & [$^{\circ}$]    & 0, 15, 30, 45, 60, 70, 90\\
          \hline
      \end{tabular}
\end{table}

\section{Method}
\label{sec:method}
We model a star surrounded by a thin disc \citep{Pringle:1981} of $N$ tracer particles, which initially move on Keplerian orbits. Each tracer or test particle represents the entire population of planetesimals in this area.
 In our study, the "disc" can also be regarded as representing all possible positions of planets of various planetary systems embedded in a debris disc. A second star perturbs the star-disc system by its flyby. Simulations were performed for varying perturber-mass to host-mass ratios $M_{21} = M_2 /M_1$ and periastron-distances $r_p$. We fix the host mass ($M_1$) to 1 \MSun\ and vary the perturber mass ($M_2$) in the range \mbox {0.3 -- 50 \MSun.}  
In this study, we only consider parabolic flybys to restrict the parameter space to a manageable size.  In low-mass clusters, parabolic flybys dominate by far \citep{Olczak:2006,Vincke:2016}. Only in high-mass clusters do hyperbolic encounters become essential. Generally, hyperbolic flybys lead to weaker effects on discs \citep{Olczak:2012,Marzani:2013,Picogna:2014}. Therefore, the results obtained here,  can also be applied in high-mass clusters but should be interpreted as upper limits of the ISO release (see discussion in Section \ref{sec:cluster}).

Discs that are old enough to have formed ISO-sized\footnote{a few tens of meters to kilometres} planetesimals from $\mu$m-sized dust grains even in the outer disc areas are usually of relatively low mass and contain very little gas \citep{Andrews:2020}. Here, low mass means that the disc mass, $m_d$, fulfils the condition,  $m_\mathrm{d}\ll 0.01~M_1$, where $M_1$ is the mass of the central star.  Viscous effects play a minor role during protoplanetary discs' late developmental stages and none at all in debris discs. ISOs are predominately released from the outer disc areas, and here viscosity is usually negligible due to the low gas content. The insignificance of self-gravity and viscosity justifies reducing the study to only account for the effects of gravitational forces by two stars on each other and the tracer particles in the discs. Therefore, the numerical treatment reduces to  $N$ three-body events \citep{Pfalzner:2005, Steinhausen:2012, Breslau:2014}.  

The initial disc size is chosen as $r_\mathrm{d}$ = 100 AU, which is characteristic for many observed dust discs. The gas disc sizes are often considerably larger \citep{Najita:2018,Andrews:2020}. Historically, in flyby simulation setups, the mass of the test particles is often kept constant, and they are distributed so that they represent the mass distribution in the disc. By contrast, in our simulations, a constant test particle distribution is chosen. Different disc-mass distributions are realized by assigning the masses to the test particles after the simulation, according to the desired density distribution in the disc \citep[for details see, ][]{Steinhausen:2012}. The advantage is a higher resolution in the disc's outer parts than in the conventional method, which is the area most relevant for ISO production. We find that in this case, a resolution with $N$~=~10 000 test particles is sufficient to determine the relative amount of planetesimals released. Besides, implementing such a flexible numerical scheme allows using one suite of simulations for any initial disc-mass distribution.

  \begin{figure}[t]
\includegraphics[width=0.48\textwidth]{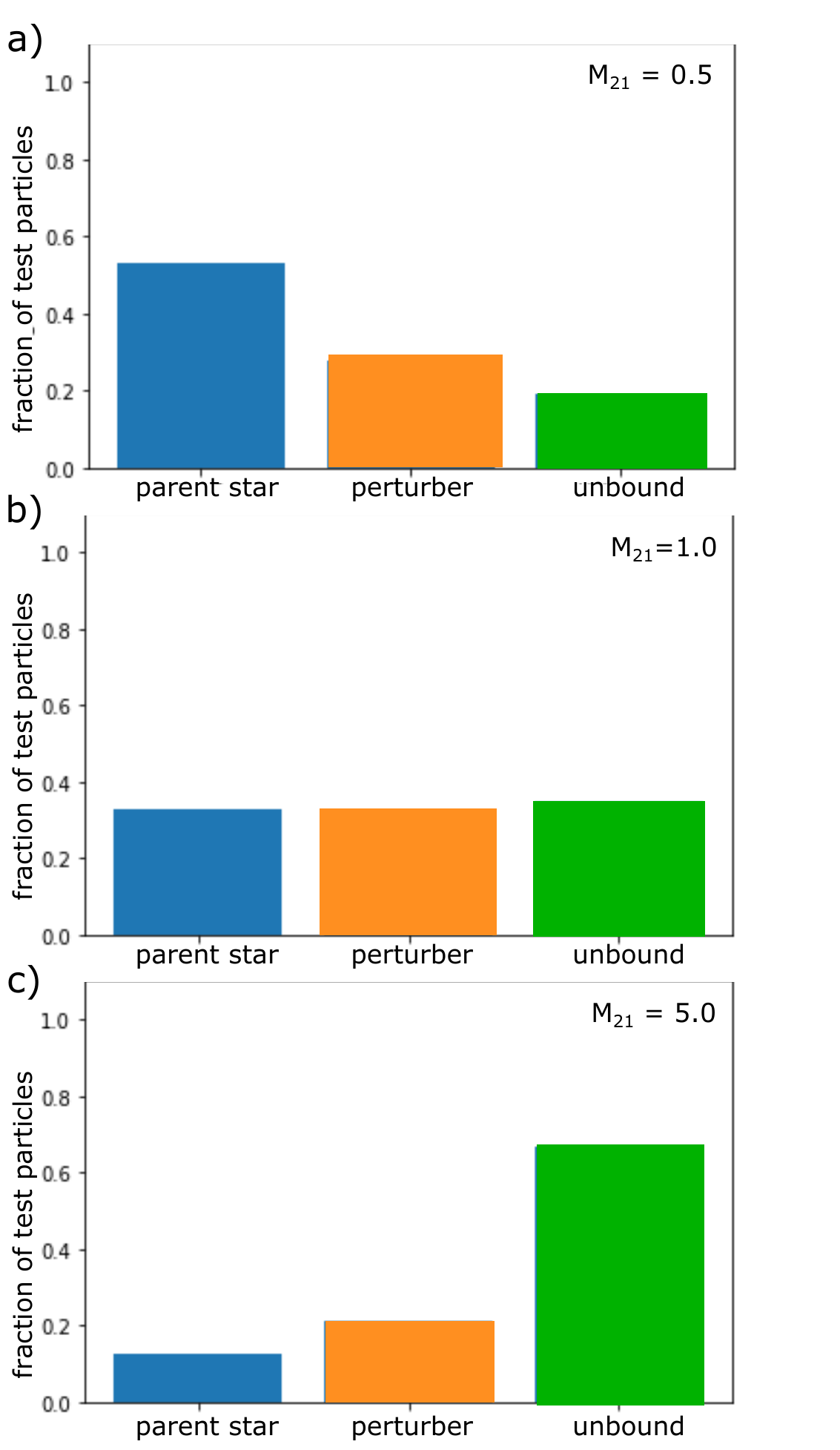}
\caption{Fraction  of test particles remaining bound to their parent star (Star 1), being captured by the perturber (Star 2) and becoming unbound during a flyby. Here, the case of a coplanar, prograde flyby at a periastron distance of 100 AU is shown. The different panels show the results for the following mass ratios between the perturber and the parent star: a)~$M_{21}$~=~0.5, b) $M_{21}$ = 1.0 and c) $M_{21}$ = 5.0.}
\label{fig:bcu}
\end{figure}
  
In principle, the perturber star itself could be surrounded by a disc, and the central star could potentially capture material from the perturber's disc \citep{Jilkova:2015}. However, this would not change the number of planetesimals becoming unbound from the central star. As we are only interested in ISO production, we assume that the perturber itself has no disc.

The tracer particles' trajectories are integrated using a Runge-Kutta Cash-Karp scheme with an adaptive time step size control. The maximum relative error between the 4th and 5th integration step is chosen to be $<$ 10$^{-7}$. 
The stars' initial positions are chosen such that the perturber's force
on the tracer particles is $<$ 0.1\% of that of the host star's. For both stars being of equal mass, this corresponds to around 40 orbits for the outermost particles.  The inner disc is usually not involved in the ISO production; therefore, we use an inner hole of 1 AU to avoid small time steps. Particles that approach one of the stars closer than 1 AU are treated as accreted. The simulations only account for the ISO production during the flyby; long-term ISO production due to scattering by the planets are not included,  as this has already been investigated, for example in  \cite{Raymond:2017}.

Here, we just gave a short overview; details of the numerical method are available in \cite{Breslau:2014, Bhandare:2016, Breslau:2017}. In section \ref{sec:cluster} we build on our previous results concerning the average disc size induced in a variety of young star clusters by stellar flybys. For details of the cluster simulations' numerical method, we refer the reader to \cite{Vincke:2016} and \cite{Vincke:2018}. For simplicity, binary flybys, although possibly significant in this context \citep{Li:2020b}, are not considered here. The reason is twofold: binaries increase the parameter space considerably and make the problem much more complex. However, our results are also applicable to close binaries with circumbinary discs. Future studies should include all types of binaries.

\begin{figure}[t]
\includegraphics[width=0.52\textwidth]{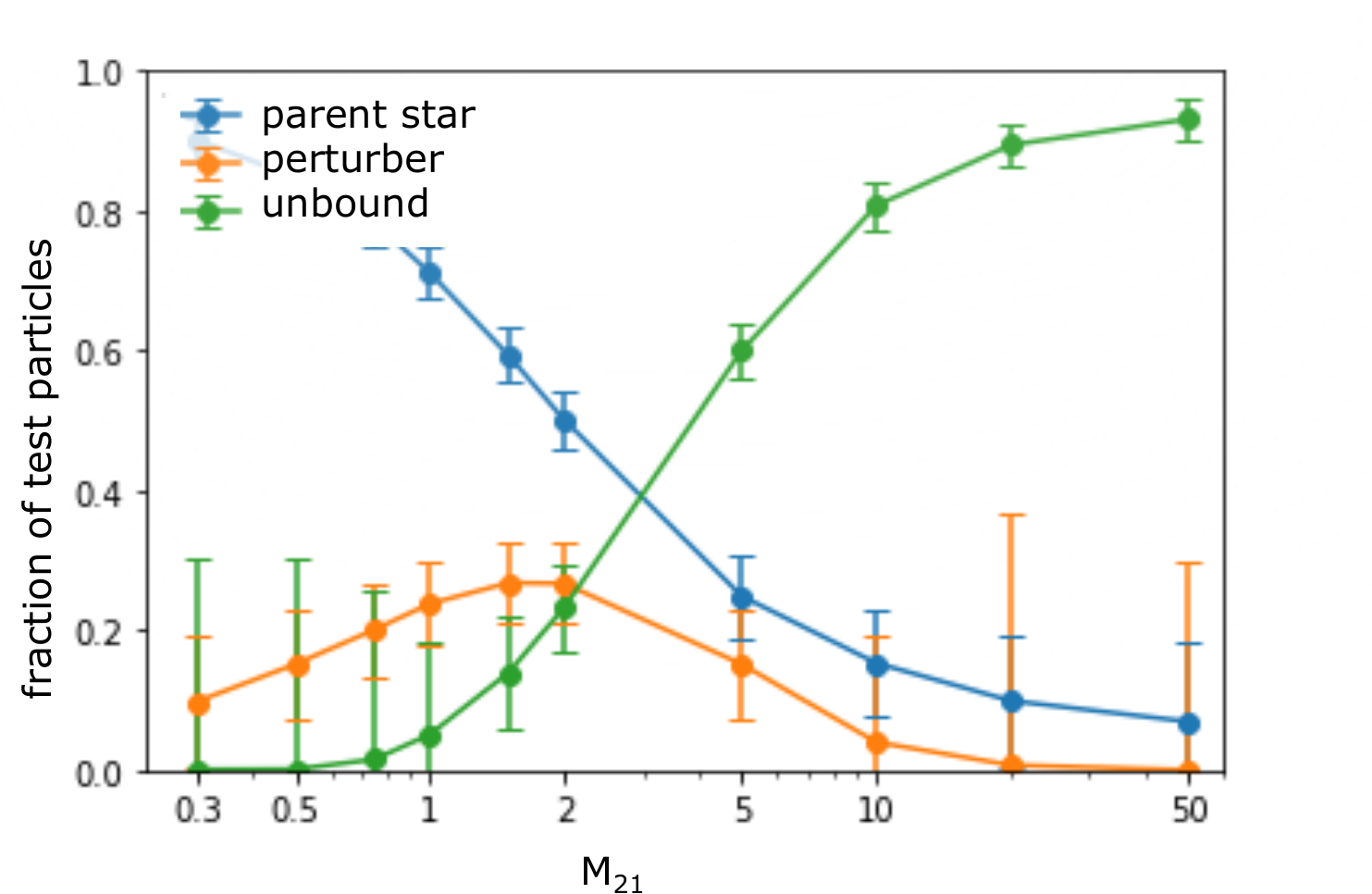}
\caption{Fraction of test particles remaining bound to the parent star (blue), being captured by the perturber (orange) and becoming unbound (green) as a function of perturber mass for a flyby with a periastron distance of 160 AU. }
\label{fig:bcu_mass}
\end{figure}

\section{Results}
\label{sec:results}

Here, we concentrate on the particles that become unbound from a disc during a close stellar flyby. First, in subsection~3.1, we discuss the prograde, coplanar case as this is the most disruptive scenario, and it already illustrates the main dependencies on perturber mass and periastron distance. Afterwards, in subsection~\ref{sec:inclination}, we demonstrate the role of inclination and angle of periastron on the effectiveness of the ejection of planetesimals. A complete list of the parameter range covered by our study can be found in \mbox{Table~\ref{tab:set-up}.}  

\subsection{Coplanar flybys}
\label{sec:results_coplanar}

A close flyby can lead to particles becoming unbound or being captured by the perturber. Central to the question of ISO production is the relative amount of planetesimals becoming unbound. In many environments, distant flybys are much more common than close flybys. In these cases, the outer fringes of a disc are perturbed at most, and a relatively small fraction of planetesimals becomes unbound.  However, what distant means in this context depends strongly on the mass of the perturber. 

Figure~\ref{fig:bcu} shows the relative portion of particles remaining bound to the parent star, being captured by the perturber and becoming unbound for three close flyby scenarios where the periastron distance of the flyby is the same, but the ratio between the mass of the perturber ($M_2$) and the parent star ($M_1$) differs ($M_{21} = M_2/M_1$= 0.5, 1.0 and 5.0). The low-mass perturber case ($M_{21}$= 0.5), illustrated in Fig.~\ref{fig:bcu}a, is characteristic for weak perturbations. Here, most particles remain bound, whereas smaller fractions of planetesimals are captured or become unbound. In this particular case, the flyby is very close; however, the flyby's effect is nevertheless quite limited due to the low mass of the perturber.  The perturber captures more material from the disc than material becoming unbound. For a higher perturber mass, $M_{21}$ = 1, but otherwise same flyby parameters, nearly equal amounts of particles remain bound, are captured or become unbound (see Fig.~\ref{fig:bcu}b).  Furthermore, for an even higher mass of the perturber ($M_{21}$ = 5), most disc particles become unbound, and only a few of them are captured or remained bound.
The above example illustrates the broad spectrum of close flyby outcomes in terms of proportions of planetesimals becoming unbound and turning into ISOs. 

Figure~\ref{fig:bcu_mass} illustrates the role of the perturber's mass for the example of flybys at a periastron distance of 160~AU.  As expected, the general trend is that the lower the mass ratio between the perturber and parent star, the more particles remain bound to the host star. A perhaps surprising finding is that at mass ratios $M_{21}\leq$~2, more particles are captured by the perturber than becoming unbound. By contrast, massive perturbers ($M_{21}\geq$~5) unleash most planetesimals from their parent star but fail to capture a significant amount of planetesimals themselves. The close flyby of a very massive star ($M_{21}\geq$~10) always leads to so many particles becoming unbound that the original disc is more or less destroyed. However, as we will see in Section \ref{sec:cluster}, such a situation is not very common, even in young stellar groups.

We have analysed the entire parameter space listed in Table~\ref{tab:set-up}. The actual ratio between particles becoming captured and unbound vary, but the maximum fraction of captured planetesimals is always at $M_{21} \approx$ 2. In the following, we concentrate on the particles that are ejected during the flyby and therefore turn into ISOs. 

\begin{figure}[t]
\includegraphics[width=0.5\textwidth]{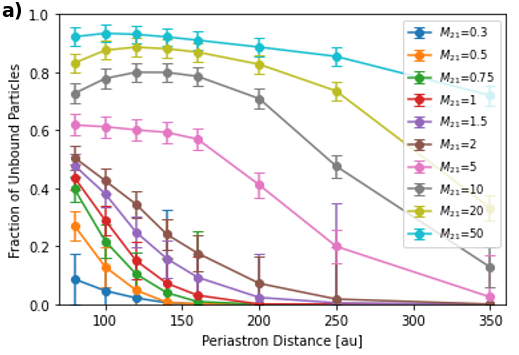}
\includegraphics[width=0.5\textwidth]{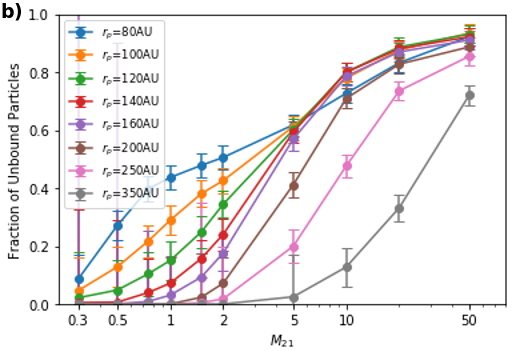}
\caption{ Fraction of unbound test particles as a function of (a) periastron distance and (b) different stellar mass ratios. }
\label{fig:Unbound}
\end{figure}

Figure~\ref{fig:Unbound}a shows the proportion of particles becoming unbound as a function of the periastron distance. The coloured lines indicate different mass ratios between the host and perturbing star. Figure~\ref{fig:Unbound}b illustrates the same but as a function of the mass ratio. In a nutshell, the stronger the interaction (larger mass of perturber or closer periastron), the larger the number of planetesimals, which become unbound from their parent star. However, both figures demonstrate the strong, often under-appreciated, role of the perturber's mass. No particles become unbound for perturber mass ratios $<$1 unless the flyby is closer than 200~AU, which corresponds to twice the disc size. By contrast, for perturber with high masses ($M_{21}>$~10), even flybys at distances farther than 300~AU can lead to more than half of the particles becoming unbound. 

However, here, one has to remember that these are the number of test particles and that we post-process these results to obtain the actual fraction of the total mass of ISOs becoming unbound. Nevertheless, when considering a $1/r$-distribution, the result is very similar as mainly the discs' outer regions are affected where the potential is relatively flat.  Table~\ref{tab:ISO} in the Appendix provides the actual values.

\begin{figure*}[t]
\includegraphics[width=18.0cm]{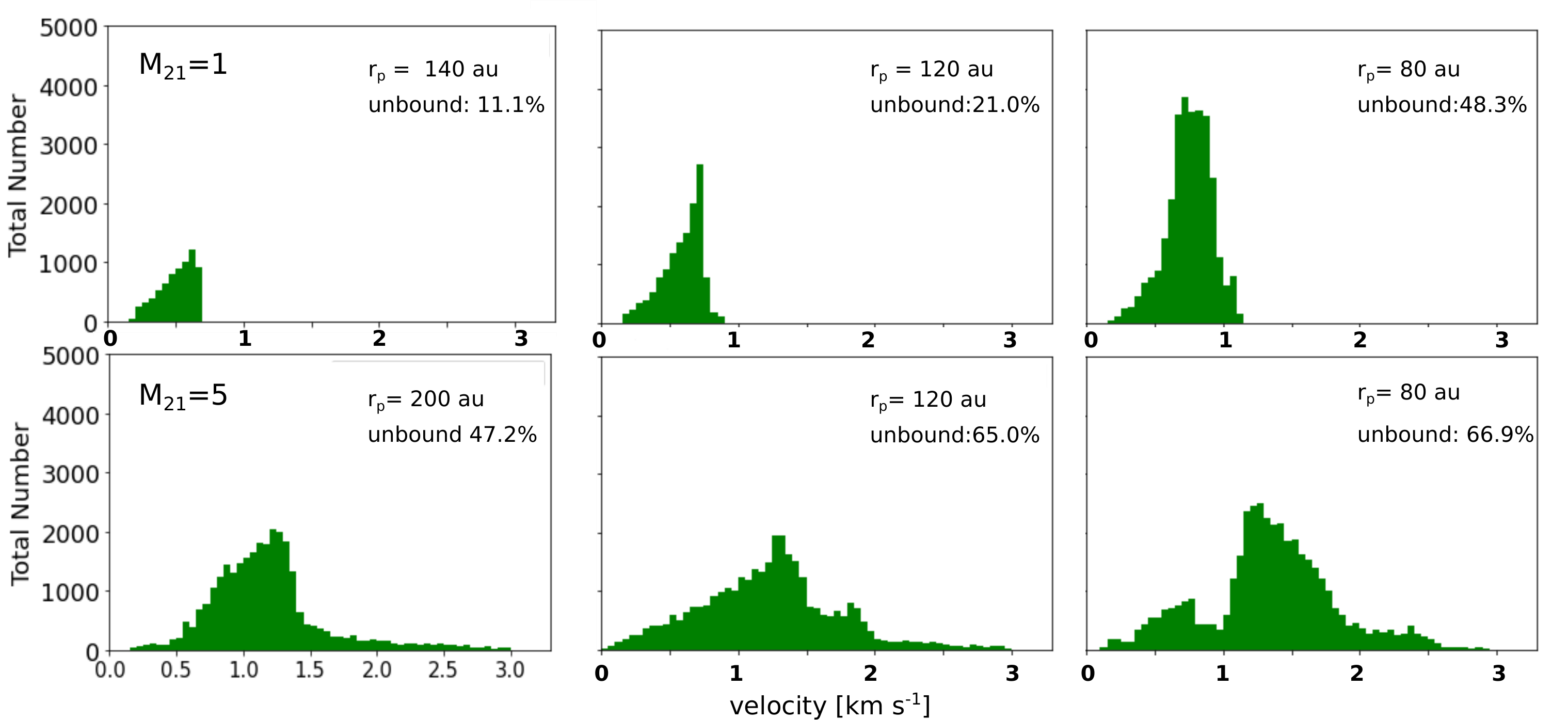}
\caption{Velocity distribution of the ISOs becoming unbound from the parent system. The top panel shows the case where perturber and parent system have the same mass, $M_{21}$ = 1, for different periastron distances. The bottom panel shows the effect of a high-mass perturber, here, exemplarily for $M_{21}$ = 5.  The velocity bin width is 0.05 km s$^{-1}$.  } 
\label{fig:velocity_distribution}
\end{figure*}

Apart from the total amount of ISOs becoming unbound, the ISOs'  ejection velocity is relevant. This ejection velocity is the ISO's velocity relative to its parent star. Figure~\ref{fig:velocity_distribution} shows the  ejection velocity distribution of the particles that become unbound for a selection of flyby scenarios\footnote{The velocity distribution of test and mass-weighted particles are very similar.}. For the case of $M_{21}$ = 1, one can see that a flyby at a periastron distance of \mbox{ $r_\mathrm{p}$ = 140 AU} leads to $\approx$~11\% of the planetesimals becoming unbound and the resulting ISOs leave the system with typical velocities of \mbox{$\approx$ 0.5 km s$^{-1}$.} The portion of planetesimals becoming unbound increases for closer flybys, and so does their mean ejection velocity. For an even closer penetrating flyby  \mbox{($r_\mathrm{p} = $ 80 AU)} more than 40\% of planetesimals become unbound, and the ejection velocity distribution peaks at 0.7~--~0.8 km s$^{-1}$. 

However, the ISO ejection velocity distribution is more sensitive to the perturber's relative mass than {\bf to} the actual periastron distance. For $M_{21}$ = 5, even a comparatively distant flyby at \mbox{$r_\mathrm{p}$ = 200 AU} results in a relatively broad ejection velocity distribution, with the peak lying at around 1.4 km s$^{-1}$. For closer flybys, the distribution becomes even broader. While the peak remains at 1.4 km s$^{-1}$, a considerable fraction of ISO leave at higher velocities of up to 2.5 km s$^{-1}$. Such a double-peaked distribution is typical for strong encounters and reflects the fact of two spiral arms developing. During such flybys always two spiral arms develop; however, they are not of equal strengths. The reason is that the first spiral arm develops as a direct reaction to the perturber star. This is where fast ISOs are produced. By contrast, the second spiral arm occurring on the opposite side of the disc forms as a reaction to the movement of the parent star \citep{Pfalzner:2003}. In distant flybys, rarely any planetesimals become unbound from this second arm. Only in strong encounters do planetesimals become unbound from the second spiral arm, despite being, on average, slower, and never exceeding velocities of 1 km s$^{-1}$. 

However, there seems to be an upper limit to the ejection velocities typically produced by stellar flybys. Even in our strongest encounter case ($M_{21}$= 50, $r_\mathrm{p}$ = 80 AU) the ISO ejection velocities stay in $>$~95\% of cases well below 3 km s$^{-1}$.

\begin{figure}[b]
\includegraphics[width=8.5cm]{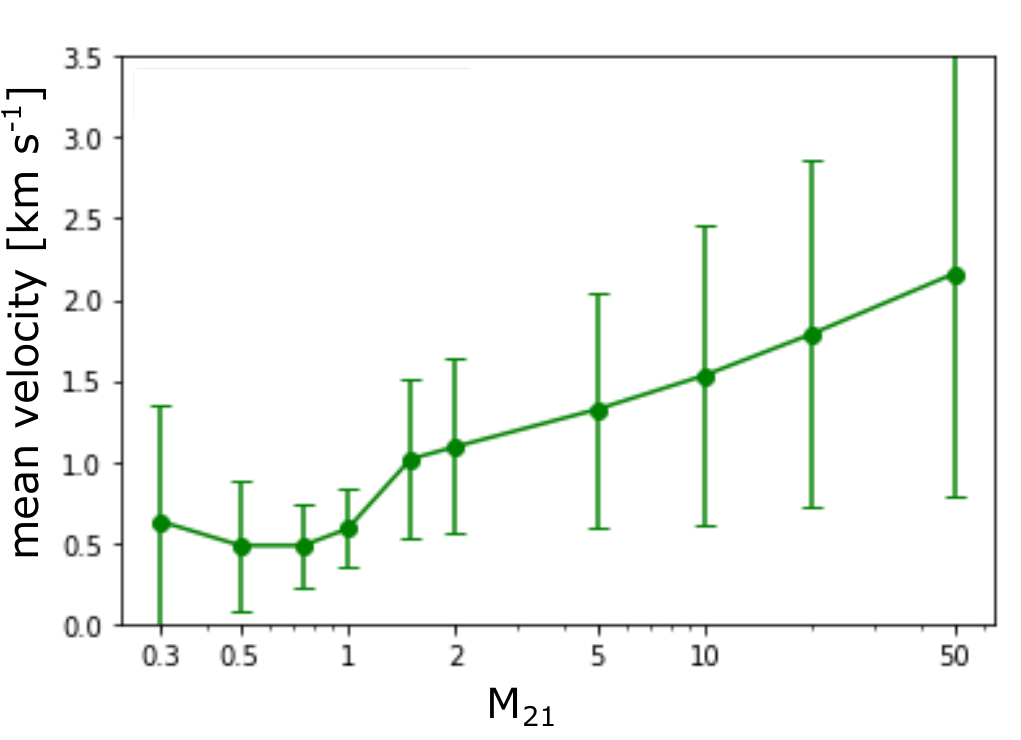}
\caption{Mean ISO velocity as a function of relative perturber mass.}
\label{fig:velocity_coplanar}
\end{figure}

The dependence of the ISOs' mean ejection velocity on the relative perturber mass is illustrated in Fig.~\ref{fig:velocity_coplanar}. This mean ejection velocity is relatively insensitive to the periastron distance; although, the actual amount of planetesimals becoming unbound depends on the periastron distance. Considering the error bars, Fig.~\ref{fig:velocity_coplanar} shows that for a relative perturber mass $M_{21}\leq$~1 the mean ISO velocity is nearly constant about 0.5 km s$^{-1}$. The mean ISO ejection velocity increases for higher velocities, with a distinct change in slope at  $M_{21}\approx$~2. As we saw earlier, this is when the particles become preferentially unbound rather than captured by the perturber.   For higher relative perturber mass, the mean ISO velocity increases nearly linearly in log space with the perturber mass; from 1 km s$^{-1}$ at $M_{21}$ = 2 to 2 km s$^{-1}$ at $M_{21}$ = 50. For additional flyby parameters, the mean ejection velocities  can be found  in Table~\ref{tab:velocities} in the Appendix.

\subsection{Non-coplanar flybys}
\label{sec:inclination}

In section~\ref{sec:results_coplanar} we concentrated on coplanar, prograde flybys. However, in real cluster environments, it is much more common that the perturber passes with some inclination ($i$) relative to the disc's plane. Previous studies have demonstrated that non-coplanar flybys are less efficient in removing material from discs than coplanar flybys of the same periastron distance and relative perturber mass  \citep{Clarke:1993, Bhandare:2016}.

\begin{figure}[t]
\includegraphics[width=0.5\textwidth]{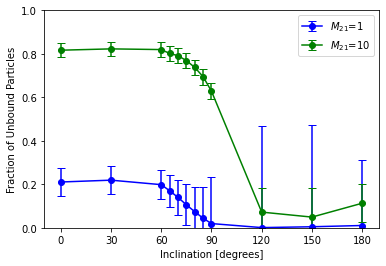}
\includegraphics[width=0.5\textwidth]{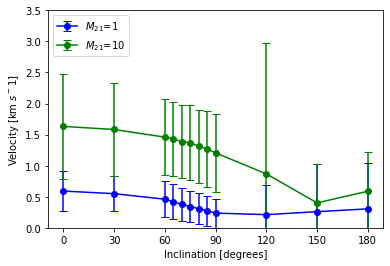}
\caption{The effect of inclination  on the ISO production. The periastron distance in all cases was $r_\mathrm{p} = $120 AU. The top panel a) shows the percentage of planetesimals becoming unbound and the bottom panel b) their mean velocity as a function of the inclination. Both properties are displayed for $M_{21}$ = 1 (blue line) and $M_{21}$ = 10 (green line). The inclination is given in degrees.}
\label{fig:inclined}
\end{figure}

Figure~\ref{fig:inclined}a shows the amount of material becoming unbound as a function of the inclination of the flybys. Here, the periastron distance of the flyby was 120 AU. For equal-mass host and perturbing stars ($M_{21}$ = 1, blue line), the amount of unbound ISOs is quite independent of the inclination, as long as the inclination remains below 60$^{\circ}$. However, for larger inclinations, flybys become increasingly inefficient in unbinding planetesimals, so that at inclinations $>$ 100$^{\circ}$ virtually no particle becomes unbound. For the case of a high-mass perturber, $M_{21}$= 10 (Figure \ref{fig:inclined}a, green line), the overall trend is similar. Here, the fraction of unbound particles is constant up to \mbox{$ i \approx$ 60$^{\circ}$} and then declines. However, roughly 10\% of the particles are still removed even in retrograde encounters.

Figure~\ref{fig:inclined}b demonstrates the dependence of the mean ejection velocity of the material becoming unbound on the inclination of the flyby. Again the inclination seems to have little effect on the mean ejection velocities of the ISOs as long as the inclination is $<$ 60$^{\circ}$. This general trend is independent of the perturber mass. For higher inclinations, the ISOs' mean ejection velocity decreases to about half of the coplanar, prograde value. Thus generally, retrograde flybys lead to lower ISO ejection velocities than prograde ones. We find that by contrast to the inclination, the angle of periastron does neither influence the number of particles becoming unbound nor the ejection velocity distribution.

The intuitive assumption that a retrograde, coplanar encounter affects the disc least does not hold here. As previous studies have already demonstrated for the disc size, flybys with an inclination in the range  110$^{\circ}$~--~160$^{\circ}$ are least efficient in removing disc material \citep{Bhandare:2016}. We see the same trend here for the material becoming unbound. 

Averaging over all inclination and angles of periastron gives a mean fraction of ISOs, $f_\mathrm{ISO}$, produced during flybys at a given periastron distance and of relative perturber mass. This mean ISO fraction is approximately half the value for the coplanar case  shown in Fig.~\ref{fig:Unbound} or Table~\ref{tab:ISO}. The mean ejection velocity of these ISOs originating from inclined flybys is approximately 70\% of the mean ejection velocity of the equivalent coplanar flyby shown in Fig.~\ref{fig:velocity_coplanar} and Table~\ref{tab:velocities}.

\cite{Hands:2019} were the only ones so far investigating the velocities of ISOs released during stellar flybys in cluster environments. In their simulations, the ISOs leave the star cluster typically with a velocity of  2.5~--~5 km s$^{-1}$. However, also velocities of up to 10 km s$^{-1}$ do occur. These values seem to differ significantly from the velocities of 0.5~--~1.5  km $^{-1}$ that we obtain here. However, this seeming discrepancy is easily solved. \cite{Hands:2019} measure the velocity relative to the cluster centre, whereas we determine the ejection velocity relative to the host star. As the cluster stars have a velocity dispersion relative to the cluster centre, the value by  \cite{Hands:2019} is the ISO velocity dispersion plus the stellar component relative to the cluster centre. In contrast, we determine the ISO ejection component relative to the star on its own.

\section{Ejection velocities in ISO formation processes}
\label{sec:other_processes}

Although there exists a clear picture of the various processes potentially creating ISOs over a star's lifetime (see Fig. \ref{fig:phases}), it is unknown which of these processes produced most of today's ISO population. Therefore, ISO properties have to be found in an ensemble sense, which will eventually statistically distinguish the different formation mechanisms. 

Here, we show that the velocities of the ISOs are one property which differs for various formation processes. Above we demonstrated that the mean velocity of ISOs released by flybys lies typically in the range 0.5~--~2 km s$^{-1}$ (see Fig.~\ref{fig:velocity_coplanar}).   In all type of cluster environments, encounters with $M_{21}\approx$ 1 are the most common scenario leading to disc mass loss and disc size reduction \citep{Olczak:2006,Vincke:2016} and therefore the release of planetesimals. Such equal-mass flybys typically produce ISOs with velocities in the range 0.6~--~0.8 km s$^{-1}$ (see Fig.~\ref{fig:velocity_coplanar}) in case of a coplanar encounter and 0.4~--~0.6 km s$^{-1}$ for non-coplanar ones. 

\begin{figure}[t]
\includegraphics[width=0.48\textwidth]{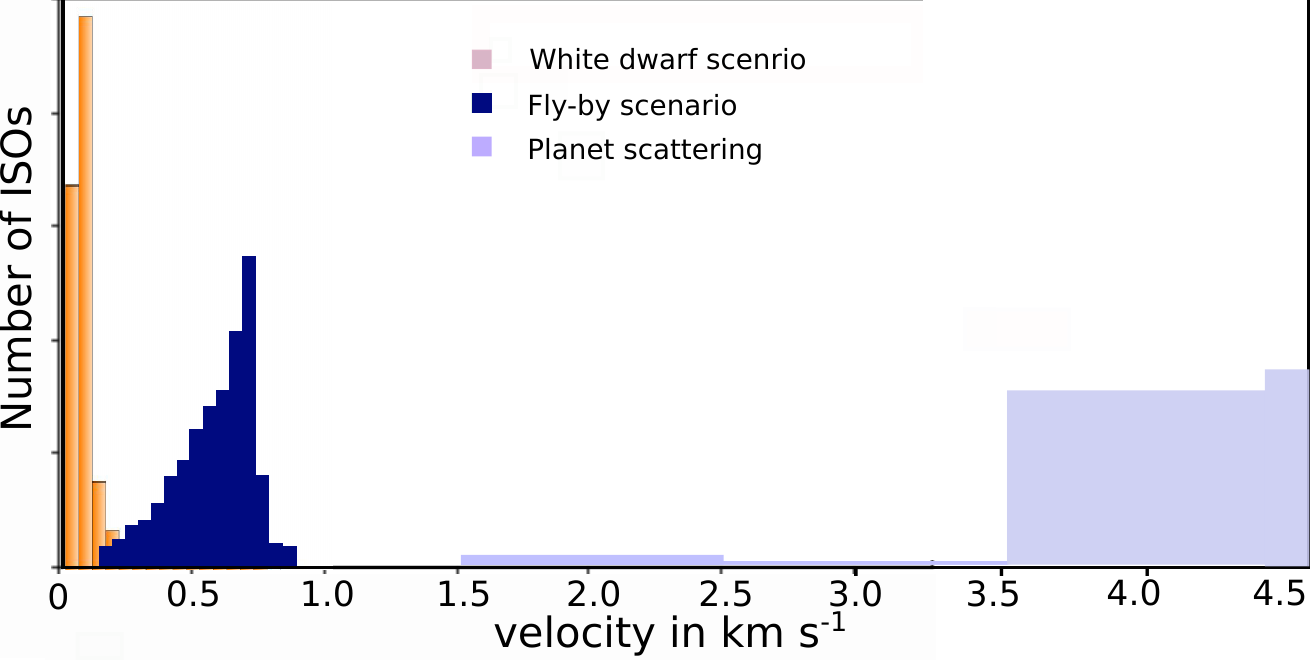}
\caption{Comparison of the ejection velocity distributions of ISOs produced during the post-main sequence phase \citep[unpublished material based on][orange]{Veras:2014vel}, the flyby scenario computed here (dark blue) and during the planet scattering phase \citep[adapted from][light blue]{Adams:2005}. The last is only partly shown, it extends further to the higher values. The velocity bin width is 0.05 km s$^{-1}$ for the flyby and post-main sequence scenario and 1 km s$^{-1}$ for the planet scattering case.}
\label{fig:comparison}
\end{figure}

In Fig.~\ref{fig:comparison} we compare the ISO velocity distributions of ISOs produced by three different formation processes --- stellar flybys (dark blue), planet scattering processes (light blue) and during the giant branch phase at the end of a star's lifetime (orange). The velocity distribution for the planet scattering process is adapted from \cite{Adams:2005}.
Their Fig. 1 shows the distribution of ejection speeds for rocky bodies on Jupiter-crossing trajectories. At least for the solar system, interaction with Jupiter leads by far to the most ejections compared to the other planets \citep{Raymond:2020}. The distribution for ISO production in the white dwarf stage resulted from the simulations described in 
\cite{Veras:2014vel}. For the flyby scenario, the velocity distribution of a coplanar flyby of an equal-mass perturber with a periastron distance of 120~AU from Fig.~\ref{fig:velocity_distribution} is shown. It should be mentioned that flybys can also trigger planetary orbital instabilities \citep{Malmberg:2011}, which in turn could produce ISOs through scattering. Although indirectly induced by a flyby, we would regard them here primarily as planet scattering process.

This comparison shows that obviously, ISOs being scattered out of their parent system due to the interaction with a planet tend to have much higher ISO velocities than those produced during fly-bys. \cite{Adams:2005} find a typical ejection speed for 4~--~8 km s$^{-1}$, compared to the 0.5~--~2 km s$^{-1}$ typical for ISOs released during flybys. In case the scattering occurs in a binary system, the ejection velocities would be even higher \citep{Adams:2005,Bailer:2018a}. These velocities would typically be 6~--~10 km s$^{-1}$ but can in extreme cases reach values above 100 km s$^{-1}$. By contrast, the ISOs produced during the giant branch phase gently drift away from their parent star at very low relative velocities of 0.1~--~0.2 km s$^{-1}$.
 Alas, each of the discussed ISO production mechanisms leads to clearly different ejection velocities.  

However, this ejection velocity is not to be confused with the derived velocity of ISOs passing through our solar system. First, the total ISO velocity consists of a superposition of the ejection velocity and the velocity of the parent star.
Consequently, the ISO velocity distribution entails the ejection velocity distribution plus the velocity distribution of the parent stars. As illustrated in Fig.~\ref{fig:phases}, ISOs are ejected by different mechanisms at different phases in a star's life. This age dependence has two consequences: First, ISOs observed today can have any age in excess of 3~--~5 Myr if produced by a stellar flyby, ISOs due to planetary scattering should be older than 10 Myr; however, ISOs produced by giant branch stars have an age of at least a few Gyr. Second, young stars have a much smaller velocity dispersion (\mbox{$<$ 10 km s$^{-1}$)} than old stars $>$ 30 km s$^{-1}$. The consequence for the velocity dispersion of ISOs (comprising ISO ejection speed plus the stellar component) is the following: just after formation, ISOs from flybys would have a mean velocity of $<$~10~km s$^{-1}$, those from planet scattering  10~--~20~km s$^{-1}$, and those from giant branch stars $>$~30~km s$^{-1}$. The entire situation is complicated by the fact that different generations of stars have produced ISOs. 

The situation is additionally complicated by the fact that the ISO's velocity might alter during their travel through interstellar space. Due to the ISO's low mass, gravitational interactions with the stars probably change the ISO velocities, but relatively rarely. Similarly, friction in the ISM will alter the ISOs velocity only to a minor extent. However, the situation might be different when ISOs pass through molecular clouds, which they pass through remarkably often \citep{Pfalzner:2020}. Another factor that might change the ISO velocities are the strong stellar tides originating from the Milky Way's spiral arms. Therefore, velocity statistics alone is probably not sufficient to reliably determine the dominant ISO's formation process. Nevertheless, the velocity statistics might be able to exclude some of them. Combining the velocity statistics with information about ISO structure will likely be the key to eventually solving this question of the dominant ISO production mechanism. 

Using velocity to distinguish between different ISO formation processes obviously requires knowledge of the velocity of more than two individual objects \citep{Zhang:2020}. The expected high detection rate of several ISOs per year for `Oumuamua-size objects by the Vera C. Rubin Observatory's Legacy Survey of Space and Time \citep[LSST,][]{Rice:2019} hopefully will allow first steps into this direction. However, the velocity differences of the processes considered above are a few km s$^{-1}$, rather than tens of km s$^{-1}$. Therefore,  probably a sample of thousands rather than hundreds of ISOs is required. The hope is that LSST will calibrate the relative mass fraction from each of these processes. However, this is not an easy task.

The velocity of an individual ISO depends on its ejection velocity, the velocity of its parent star and possible velocity alterations during the ISO's journey through the ISM \citep{Hallatt:2020,Pfalzner:2020}. Therefore, any answer to the question of whether \Ou\ and Borisov were likely released from their parent systems by a stellar flyby or another process has to remain at the level of an educated guess.  `Oumuamua had a heliocentric velocity of 26.17~km s$^{-1}$
\citep{Meech:2017b}, which corresponds to only 10~km s$^{-1}$ relative to the Local Standard of Rest \citep{Schoenrich:2010}. Given, that the velocity dispersion of stars near the Sun is between 25 and 40 km s$^{-1}$ \citep{Rix:2013}, `Oumuamua was likely ejected from its host system relatively recently and locally \citep{Meech:2017a}. Concerning the ejection mechanisms, this low velocity makes it very unlikely that `Oumuamua or Borisov were ejected by any mechanisms with a mean velocity above 5~--~10~km s$^{-1}$. This low velocity makes it unlikely that `Oumuamua was ejected by planet scattering form a binary system, or originated from a post-main-sequence star. Planet scattering leads to too high ejection speeds, whereas the age of resulting white dwarfs also means a high stellar velocity dispersion. Borisov had a higher velocity; therefore, at least planet scattering or a binary origin \citep{Bailer:2020} cannot be excluded as formation process. Equally well, an ISO released 1 or 2 Gyr ago by a stellar flyby would be an option. Just based on the velocities alone, both ISOs could have been freed from their parent star by a close stellar flyby. 

However, `Oumuamua and Borisov differed considerably in their structural parameters, with `Oumuamua being more asteroid-like, while Borisov was very much like a solar-system comet. Asteroids and comets are spatially well separated in the solar system, with volatile-poor asteroids residing in the inner solar system (distance to the Sun $<$~5~AU). In contrast, volatile-rich comets orbit most of their time outside 30 AU. Our results show that flybys very rarely release planetesimals from within 30~AU. The exception might be dense stellar clusters, similar to Arches or Westerlund 2 (see section~\ref{sec:cluster}), where even complete disc destruction can occur for up to 40\% of stars \citep{Olczak:2012,Vincke:2016}. However, there are just a few environments that might have been so dense in their past in the solar system's close vicinity.  The situation is different for the comet-like Borisov, which a stellar flyby might have well produced. 

Nevertheless, it would be premature to conclude that different ejection mechanisms produced `Oumuamua and Borisov. So far, we know little about the processes that affect ISOs on their journey through the ISM \citep{Zhang:2020}. What seems inevitable is that a considerable fraction should pass through molecular clouds \citep{Pfalzner:2020}, were the gas and dust density is much higher than in the ISM itself. It might be that the structural differences in the two objects are somewhat due to their diverse history on their journey from their parent star.

Here, we concentrated on three ISO production mechanisms. Other ISO production mechanisms have also been suggested; for example, the ejection during the planet instability phase \citep{Raymond:2017}, ejection from binary systems \citep{Smullen:2016,Wyatt:2017, Jackson:2018,Cuk:2018} and drifting from the outer Oort cloud \citep{Brasser:2010,Kaib:2011,Hanse:2018}. However, these studies mainly concentrated on the amount of planetesimals released rather than their typical velocity. It can be expected that the ejection speed during planet instability could be similar to that during planet scattering, which would put them in the range of 4~--~8 km s$^{-1}$. The drifting from the Oort cloud likely happens at relatively low velocities of $\ll$ 0.3 km s$^{-1}$. By contrast,  binary systems likely eject planetesimals at very high velocities ($\gg$ 15 km s$^{-1}$). The complete ISO velocity distribution contains multiple components reflecting the parent system's different ages and various ejection speeds based on ISO production processes' diversity. Disentangling the various components will be one of the significant challenges in the future.

\section{Different cluster environments}
\label{sec:cluster}
Here, we estimate how efficient different cluster environments are in producing ISOs. We will base these estimates on the average truncation disc size resulting from close flybys in various cluster environments \citep{Vincke:2016,Vincke:2018}.  These results were obtained by performing N-body simulations for various types of stellar groups following all stars' trajectories.  These simulations also included a realistic initial mass function for the stellar mass distribution in the clusters. The parameters (mass ratio, periastron distance, eccentricity) of all close flybys occurring over the first 10 Myr of a young cluster's evolution were recorded. Several realisations of the same cluster parameters were performed to ensure that the obtained results on the occurrence rate of specific flyby parameters were valid to an uncertainty of 3\%.  A short summary of the effect of the cluster dynamics on discs and planets is given in the next paragraph; for more details on this topic in general, see \citet[][ and references therein]{Vincke:2016,Vincke:2018}. 

Young clusters fall basically into two groups. The so-called associations are bound in the gas-embedded phase but largely dissolve after gas expulsion at the end of the star formation process. By contrast, open clusters are much more compact and can remain bound after gas expulsion for hundreds of Myr, if not Gyr. The open clusters' compact nature means a higher stellar density, more close flybys and a smaller mean disc size.  Figure~\ref{fig:cluster}a illustrates this by showing the mean truncation disc size as a function of the cluster type. Blue bars indicate open clusters, red bars associations, both for a different number of stars. Each is given for different cluster masses \citep[for actual mass values see][]{Vincke:2018}. Names of example clusters that fall in the different categories are added. Basically, in all long-lived open clusters, close stellar flybys lead to mean disc sizes below 100~AU. Due to their longevity, these clusters are prone to close encounters beyond the first 10~Myr.  Consequently, most stars in open clusters can be expected to populate the ISM with large quantities of ISOs produced by close flybys.

\begin{figure*}[t]
\includegraphics[width=0.9
\textwidth]{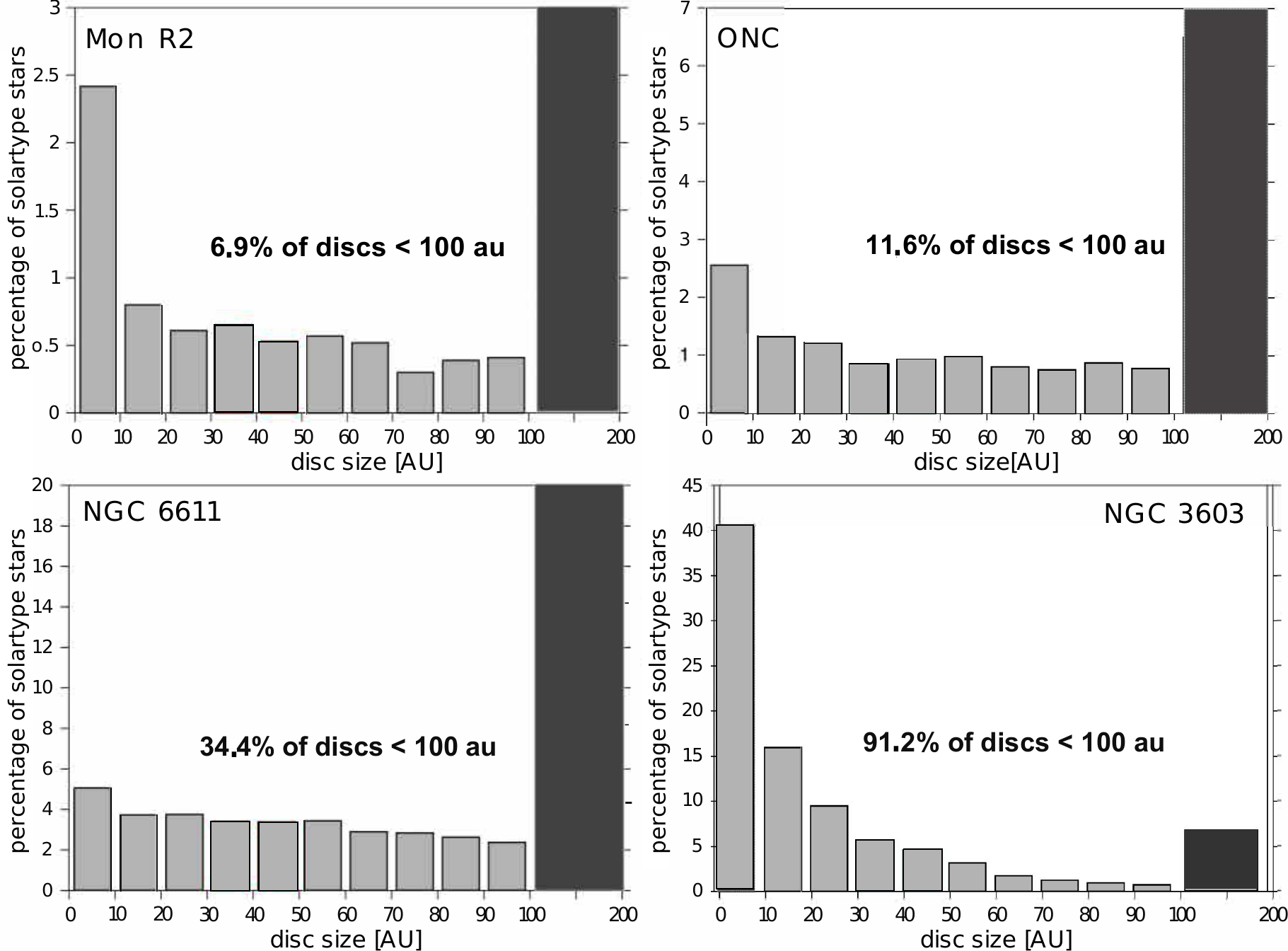}
\caption{System size distribution of simulated clusters representative of Mon R2, ONC, NGC 6611, and NGC 3603. The dark gray bar contains all disks having final sizes between 100 AU $\leq R_{d} <$ 200 AU. The fractions shown are those for solar-type stars, meaning those in the mass range \mbox{0.8 \MSun\ $ < M_1  <$ 1.2 \MSun\ .} Adapted from \cite{Pfalzner:2020}, Fig. 4.}
\label{fig:cluster_size_distribution}
\end{figure*}

\begin{figure}[t]
\includegraphics[width=0.48
\textwidth]{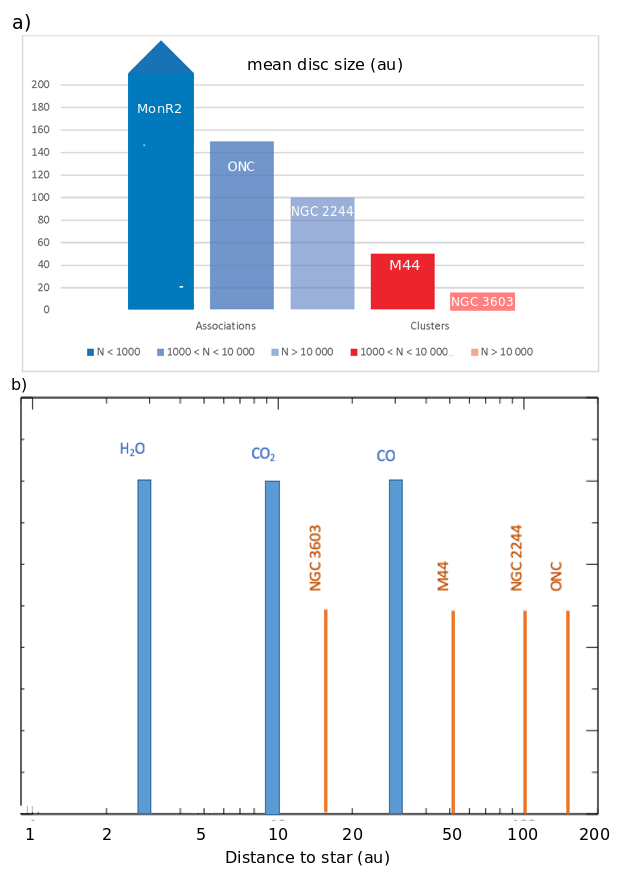}
\caption{a) shows the disc truncation radii in different cluster environments. Blue bars indicate associations, red bars open clusters of different mass \citep[adapted from][]{Pfalzner:2019}. b) shows a comparison of the different snow lines (blue) and the mean different disc truncation radii of different cluster environments (orange) from (a) as a function of the distance to a central star.   }
\label{fig:cluster}
\end{figure}

The situation is different for associations, where only in massive groups with $N >$ 10 000 most stars experience flybys closer than 100~AU. Here, $N$ stands for the number of stars constituting the association. In lower-mass clusters, only a fraction of stars is releasing ISOs due to close flybys. These are the ones harbouring extended discs ($r_d$~>~100~AU) and the few ones experiencing close flybys. The fraction of stars experiencing close flybys that lead to small discs depends strongly on the cluster environment (see Fig. \ref{fig:cluster_size_distribution}).  Consequently, the number of ISOs produced per cluster is probably considerably higher for massive open clusters than for low-mass associations. However, considering that only 10\% of all stellar groups develop into long-lived clusters, it remains open which type of cluster produces most ISO in total.

There are two caveats with this approach; first, in these simulations, most flybys occur within the first 3~Myr of cluster development, and it is not clear to what sizes planetesimals have grown at that point. Second, flybys that lead to smaller than mean truncation disc sizes might contribute disproportionately much to ISO production so that the values obtained here underestimate the total ISO production of a star cluster.

Unlike low-mass associations, massive open clusters likely release at least two types of ISOs. The reason is the location of the ice lines in discs. The ice line is the particular distance from the central star, where the temperature is sufficiently low for volatile compounds to condense into solid ice grains. Beyond the ice lines, many more solid grains of a specific variety are available for accretion onto planetesimals. Typical volatiles condensing into such solid ice grains are water, ammonia, methane, carbon dioxide, and carbon monoxide.   As different volatiles have different condensation temperatures, the location of their respective ice lines differ. The ice lines' distance is sensitive to the physical model used, the theoretical disc model and the mass of the parent star \citep{Habayashi:1981,Podolak:2004,Martin:2012,Dangelo:2015}. As the disc evolves, the radial position of the condensation/evaporation front changes over time \citep{Owen:2020}. However, here, we use the current day values in the solar system for simplicity. They are $\approx$~3~AU for the H$_2$O, $\approx$~10~AU for the CO$_2$ and $\approx$~30~AU for the CO \citep{Qi:2013,Musiolik:2016}. 

In Fig.~\ref{fig:cluster}b the location of these ice lines (blue bars) is compared to the mean truncation disc size (orange bars) in the diverse cluster environments shown in Fig.~\ref{fig:cluster}a. The figure demonstrates clearly that the mean truncation disc sizes are in most environments outside of the CO ice line so that it can be expected that most ISOs produced by stellar flybys will show no deficiency in CO or CO$_2$. However, the situation is different for very massive dense, long-lived clusters. Here, more than half of the stars have discs truncated to less than 30~AU, meaning they are likely depleted in CO. This means a depletion in CO could be an indicator for the type of parent cluster.

However, even in low-mass associations, a small fraction of stars experience close flybys. For example, even in relatively sparse associations like Mon R2, $\approx$ 4\% of the discs are truncated to less than 30~AU and $\approx$ 2\% to less than 10~AU (see Fig. \ref{fig:cluster_size_distribution}). Therefore, a relatively small number of stars is responsible for most of the ISO production in such environments.  It also means that even these low-mass clusters can produce a small number of CO- and CO$_2$-depleted ISOs. However, in dense clusters like NGC 3603, up to 65\%  of the produced ISOs would be depleted in CO and $\approx$ 40\% in CO$_2$. Close to 20\% would also be deficient in H$_2$O. These ISOs would be more asteroid- rather than comet-like. Therefore, a depletion in CO could be an indicator of the type of parent cluster. 

Next, we want to give a rough estimate of how much matter in the form of ISOs is typically produced per star in clusters like the ONC. Here, we assume that only flybys leading to disc sizes $<$~100~AU produce ISOs.  Again using Fig. \ref{fig:cluster_size_distribution} as a guideline, only $\approx$ 10\% of stars in the ONC experience flybys that lead to such truncation disc sizes. Assuming that most of these flybys are between equal-mass stars and using the relation by \cite{Breslau:2014} for the resulting disc size, 
\begin{equation}
	R_{\mathrm{d}} = 1.6 \cdot r_\mathrm{p}^{0.72} \cdot M_{21}^{-0.2}, 
	\label{eq:breslau_disc_size}
      \end{equation}
we can deduce the periastron distance that is necessary to produce such a disc size. For $M_{21}$=1, Eq.~\ref{eq:breslau_disc_size} reduces to $r_{\text{peri}}~\approx~0.52~\times~R_{\text{d}}^{1.4}$. Summing over the disc size distribution, and averaging this amount over the population of cluster stars, leads to the conclusion that 2.5\% of a disc's mass would be ejected per star on average. 
This value agrees with the 1~--~4\% of the disc mass that \cite{Hands:2019} find in their simulations. A conservative estimate of the total disc mass before truncation would be $m_d =$~0.0001~\MSun \citep{Williams:2017}. Therefore, at least 0.8 Earth masses of ISOs are ejected per star in ONC-like environments. This value could easily increase by a factor of ten for higher initial disc masses.

The ISO production per {\bf star} is much higher in dense clusters like NGC 3603 or the Arches cluster than for the ONC. Applying the above procedure to such dense clusters, the ISO production rate amounts to $>$ 50\% of the entire disc mass, which corresponds to $\approx$ 30~--~60 Earth masses per star.  However, the assumption of parabolic flybys, still valid for environments like the ONC \citep{Olczak:2006}, breaks down for such compact, high-mass clusters  \citep{Olczak:2012}. For these high-mass clusters, our results overestimate the actual ISO release by a factor of 2~--~3. Considering this,  the average ISO decreases to 10\%~--~20\% of the disc mass, equivalent to 8~--~16 Earth masses.

Newly liberated ISOs may leave the star cluster and immediately become free-floating planetesimals, or stay inside the cluster until the cluster itself dissolves. For the long-lived clusters, this dissolution may take several 100~Myrs or even Gyr. By contrast, low-mass associations typically dissolve on a 10~Myr timescale. Thus, very young free-floating ISOs (< 10 Myr) produced by close stellar flybys are likely to originate from associations rather than long-lived clusters. 

The given numbers are only a first estimate and should be tested in dedicated cluster dynamics investigations. However, the approach presented here should suffice for a rough order of magnitude analysis.  Besides, the question of the importance of the ISO release by flybys relative to other ISO production mechanisms is intricately linked to the typical star formation environment. However, it is still an open question whether most stars form in the many associations containing only a couple of stars or the few massive star clusters containing several ten thousands of stars. Therefore, future statistical approaches built on velocity and composition are probably better measures to determine the relative importance of flybys as an ISO production mechanism than numerical N-body simulations alone.

\section{ISO production by the Solar System}
\label{sec:Sun}

We live in a planetary system that likely produced many ISOs by the various processes discussed in Section~\ref{sec:intro}. A past close flyby likely produced a considerable proportion of these solar interstellar objects (SISOs).  In the early life of our Sun, close flybys were much more common than nowadays \citep{Bailer:2018b}. A strong indicator for a past close flyby is the relatively sharp drop in the solar system's mass content at about 30~AU \citep{Morbi:2004,Pfalzner:2020}. Beyond this outer edge of the solar system, only the equivalent of  0.061  times the Earth's mass is contained in total in the Kuiper belt objects \citep{DiRuscio:2020}. Likely, the solar system's disc was initially considerably more extended and truncated either in the very early stages or at later times. Several processes have been suggested as a cause for the solar systems' disc truncation: close stellar flybys  \citep{Ida:2000,Kenyon:2004,Spurzem:2009,Pfalzner:2018,Batygin:2020}, external photo-evaporation by nearby massive stars \citep{Adams:2006,Owen:2010,Mitchell:2011} and a nearby supernova explosion \citep{Chevalier:2000}. Although all these processes lead to smaller disc sizes, it is unclear whether external photo-evaporation or a nearby supernova would lead to the observed sharp outer edge. By contrast, close flybys more or less inevitably lead to a distinctive sharp outer edge \citep{Breslau:2014}.

Using the above results, we can estimate the mass of SISOs. The outer edge location is one constraint that narrows down the flyby's parameters. Additional information comes from the trans-Neptunian objects (TNOs) that restrain the parameters of a possible flyby. Unlike the planets themselves, the TNOs do not orbit on coplanar, circular orbits around the Sun, but move mostly on inclined, eccentric orbits and are distributed in a complicated way. A close stellar flyby can explain the TNO orbits, including those of the extreme TNOs like Sedna \citep{Pfalzner:2018}. These constraints can be used to derive limits on the parameters of the solar system forming flyby. They find that a prograde, parabolic flyby of a star with $M_2$~=~0.5~\MSun, a perihelion distance of 100~AU at an inclination of 60$^{\circ}$, tilted by 90$^{\circ}$ gives a good fit to the solar system properties.

Here, we use these parameters to estimate the mass of solar ISOs produced during such a flyby. We assume an initial disc size of 100~AU. Section~\ref{sec:inclination} showed that such an inclination and angle of periastron reduces the SISO production to 80~--~90\% of that of a coplanar flyby. Therefore, for such an inclined flyby  10~--~15\% of the total disc mass is released. As the solar system profile is steeper than the $1/r$ profile assumed here, the total mass of SISOs produced would have been $\approx$~5\%~--~8\% of the disc mass. Comparing this value with the per star production in an ONC-like cluster (see section~\ref{sec:cluster}), the Sun produced a relatively large amount of ISOs during this flyby. If we make a conservative estimate of the disc mass of 0.0001~\MSun\  again
\citep{Williams:2017}, the total mass of all ISOs ejected during the solar system shaping flyby corresponds to at least 2~--~3 Earth masses. These released ISOs would have left the solar system at mean velocities of $\approx$~0.7~km s$^{-1}$.

\section{Summary and Conclusion}

This study focused on the production of ISOs by close stellar flybys typically occurring in dense young star clusters. We investigated the quantity and velocities of the planetesimals that become released by such flybys. 
The results of this study can be summarized as follows:

\begin{itemize}
\item  We have tested how the planetesimal release depends on the specific flyby parameters. Generally, the more massive and closer the flyby, the more planetesimals are ejected and become ISOs. Similarly, planetesimals become most easily unbound in coplanar, prograde flybys, whereas strongly inclined flybys ($>~$70$^\circ$) lead to considerably smaller amounts of ISOs which mostly vanishes for retrograde flybys. The actual dependencies on the flyby parameters have been quantified in Fig.~\ref{fig:Unbound} and Table~\ref{tab:ISO}.
\item However, not all particles that become unbound from the parent star turn automatically into ISOs.  Some fraction of particles is also captured by the perturber. For specific flybys, the proportion of captured planetesimals can exceed those becoming free-floating ISOs. The number of particles that become bound/accreted by the perturber peaks when the perturber mass is 1.5~--~2.0~\MSun, independent of the periastron distance.
\item The mean velocity of ISOs increases for higher perturber masses, is relatively independent of periastron distance and decreases as a function of inclination.
\item The velocity distribution of unbound planetesimals for typical perturbers ($M_{21} <$~1.5) ranges from 0.2 to 1.5~km s$^{-1}$.  For high-mass perturbers ($M_{21} >$~5) the ISOs' escape velocities can increase to 3.0~km s$^{-1}$.
In this case, the velocity distributions show a distinguishable "double peak" feature reflecting the characteristic two spiral arms produced by flybys.
\item These ISO ejection velocities differ from those of other ISO production processes. Specifically, it is smaller than the velocities induced by the planet scattering processes (4~--~8~km s$^{-1}$). At the same time, they are faster than those produced towards the end of a star's lifetime;  those are usually below $\ll$~0.3~km s$^{-1}$). Therefore, the ISO velocity is an essential parameter for restraining the dominant ISO production mechanism.
\item In most cluster environments, it is a small subset of the stars that produce most of the ISOs. Nevertheless, the equivalent of 0.85 Earth-masses of ISOs is produced per star in ONC-like environments. The mass of ISOs released from their parent star can be considerably higher in denser stellar clusters like NGC 3603.
\item The solar system likely released the equivalent $\approx$~5~--~8\% of the disc mass, which for a disc of mass 0.0001~\MSun\ corresponds to 2~--~3 Earth masses of ISOs. These solar ISOs left the Sun with a mean velocity of $\approx$~0.7~km s$^{-1}$.
\end{itemize}

This study is only a first step for understanding the release of planetesimals due to close stellar flybys. Significantly, the cluster dynamics itself and the ISO dynamics in the cluster have to be better understood in the future. In particular, the timescales on which planetesimals form in the outer disc areas have to be better constrained. Nevertheless, this work might serve as a stepping stone for future studies on star-disk encounters as sources for ISOs such as `Oumuamua and Borisov.\\

\begin{acknowledgements}
We thank the referee, who has provided helpful comments which have improved the manuscript.
DV gratefully acknowledges the support of the STFC via an Ernest Rutherford Fellowship (grant ST/P003850/1).
\end{acknowledgements}

\bibliographystyle{aa} 
\bibliography{references} 

\begin{appendix}

\section{Portion of ISOs produced by a flyby}

\begin{table}[h]
\centering   
    \caption{Percentage of unbound particles as a function of periastron distance and perturber mass ratio. Here, the case of coplanar prograde flybys is shown.}
    \label{tab:ISO}
    \begin{tabular}{l|llllllll} 
    \hline \hline
    \multicolumn{1}{l|}{$M_{12}$} & \multicolumn{8}{c}{$r_p$ [AU]} \\
    \hline 
                \noalign{\smallskip}
             &80  & 100  & 120  & 140  & 160  & 200  & 250  & 350  \\
      \hline
             \noalign{\smallskip}     
      
      0.3 & 0.122 & 0.071 & 0.036 & 0.008 & 0.008 & --- & --- & --- \\
      
      0.5 & 0.346 & 0.192 & 0.080 & 0.011 & 0.001 & --- & --- & --- \\
      
      0.75 & 0.451 & 0.289 & 0.157 & 0.064  & 0.015 & --- & --- & --- \\
      
      1 & 0.482 & 0.349 & 0.209 & 0.111 & 0.051 & --- & --- & --- \\
      
      1.5 & 0.522 & 0.429 & 0.311 & 0.213 & 0.139 & 0.039 & 0.007 & --- \\
      
      2 & 0.549 & 0.471 & 0.396 & 0.303 & 0.232 & 0.107 & 0.000 & --- \\
      
      5 & 0.669 & 0.664 & 0.646 & 0.626 & 0.600 & 0.472 & 0.279 & 0.042\\
      
      10 & 0.781 & 0.806 & 0.817 & 0.817 & 0.807 & 0.747 & 0.567 & 0.197\\
      
      20 & 0.861 & 0.886 & 0.898 & 0.899 & 0.893 & 0.860 & 0.781 & 0.442\\
      
      50 & 0.931 & 0.943 & 0.944 & 0.939 & 0.931 & 0.913 & 0.886 & 0.772\\
    \hline  
    \end{tabular}
\end{table}

\section{Velocities of ISOs produced by a flyby}
\begin{table}[h]
Table \ref{tab:velocities} gives the mean velocity of the particles becoming ISOs for different flyby parameters. Naturally, meaningful mean ISO velocities can be derived if the number of ISOs is sufficiently high. Therefore, the values for $M_2/M_1$ = 0.3 and the values for high periastron values have a much
larger error than the other values.\\

  \centering
    \caption{Mean velocity of unbound particles as a function of periastron distance and perturber mass ratio. Here, the case of coplanar prograde flybys is shown. }
    \label{tab:velocities}
        \begin{tabular}{l|llllllll} 
    \hline \hline
        \multicolumn{1}{l|}{$M_{12}$} & \multicolumn{8}{c}{$r_p$ [AU]} \\
    \hline 
                \noalign{\smallskip}
             &80  & 100  & 120  & 140  & 160  & 200  & 250  & 350  \\
 
      \hline
                  \noalign{\smallskip}
      0.3 & 0.751 & 0.729 & 0.683 & 0.550 & 0.434 & --- & --- & --- \\
      
      1 & 0.749 & 0.670 & 0.591 & 0.507 & 0.411 & --- & --- & --- \\
      
      2 & 1.053 & 1.068 & 1.040 & 1.065 & 1.057 & 1.028 & 1.327 & --- \\
      
      5 & 1.385 & 1.434 & 1.298 & 1.429 & 1.357 & 1.248 & 1.105 & 0.998\\
      
      10 & 1.594 & 1.626 & 1.479 & 1.603 & 1.573 & 1.473 & 1.380 & 1.168\\
      
      50 & 2.004 & 2.049 & 2.261 & 2.152 & 2.205 & 2.220 & 2.233 & 2.073\\
    \hline  
    \end{tabular}
\end{table}

\end{appendix}

\end{document}